\documentclass[final,numbers,sort&compress]{aipproc}
\usepackage{amsmath}
\usepackage{graphicx}
\graphicspath{{./Figures/}}
\layoutstyle{6x9} 

\newcommand{\trace}{\mbox{Tr}}                                     
\renewcommand{\Re}{\textrm{Re}\,}                                  

\newcommand{\Fig}[1]{Fig.\,\ref{#1}}

\newcommand{\atconference}{~at \emph{Quark Confinement and the Hadron
                            Spectrum VI}, Villasimius, Sardinia,
                            Italy, September 21-25, 2004}

\newcommand{\preprintnumbers}{\thispagestyle{empty} \vspace{-25mm}
            \begin{flushright}
              \normalfont \small HU--EP--04/68, LU-ITP 2004/044\\
                          \small December 2004
            \end{flushright}                           
            \vspace{8mm}}    

\begin{document}

\title{%
\preprintnumbers    
The influence of Gribov copies on the \linebreak[1] gluon and ghost
propagator \footnote{Talk presented by A.~Sternbeck%
\atconference       
.}}

\classification{11.15.Ha, 12.38.Gc, 12.38.Aw} 
\keywords{ghost and gluon propagator, Gribov problem, Faddeev-Popov 
          operator eigenvalues}

\newcommand{\huaddress}{Institut f\"ur Physik, Humboldt-Universit\"at
                        zu Berlin, D-12489 Berlin, Germany}
\newcommand{\leipzig}{Universit\"at Leipzig, Institut f\"ur
                      Theoretische Physik, D-04109 Leipzig, Germany}

\author{A.~Sternbeck}{address=\huaddress}
\author{E.-M.~Ilgenfritz}{address=\huaddress}
\author{M.~M\"uller-Preussker}{address=\huaddress}
\author{A.~Schiller}{address=\leipzig}

\begin{abstract}
  The dependence of the gluon and ghost propagator in pure $SU(3)$
  gauge theory on the choice of Gribov copies in Landau gauge is
  studied. Simulations were performed on several lattice sizes at
  $\beta=5.8$, $6.0$ and $6.2$. In the infrared region the ghost
  propagator turns out to depend on the choice, while the impact on the gluon
  propagator is not resolvable. Also the eigenvalue distribution of the 
  Faddeev-Popov operator is sensitive to Gribov copies.
\end{abstract}

\maketitle


Studying non-pertubative features of QCD such as confinement, there
are two common approaches: lattice gauge theory and Dyson-Schwinger
equations. From the latter approach there are promising results in
recent years \cite{Alkofer:2000wg} about the infrared behavior of the gluon $D$
and the ghost propagator $G$. Denoting by $Z$ the dressing functions of
the corresponding propagator, in Landau gauge they can be written as
\begin{equation}
  D_{\mu\nu}(q^2) = \left(\delta_{\mu\nu} - \frac{q_{\mu}q_{\nu}}{q^2}
                    \right) \frac{Z_{gl}(q^2)}{q^2}\quad\textrm{and}\quad 
  G(q^2) = \frac{Z_{gh}(q^2)}{q^2}\;.
\end{equation}
According to \cite{Alkofer:2000wg} in the low-momentum region the dressing
functions are proposed to behave as $~Z_{gl}\propto(q^2)^{2\kappa}~$ and
$~Z_{gh}\propto(q^2)^{-\kappa}~$ with a common value \mbox{$\kappa\in(0.5,1)$}.
The infrared suppression of the gluon propagator and the enhancement
of the ghost propagator at low-momentum is in agreement with the
Zwanziger-Gribov horizon condition 
\cite{Zwanziger:2003cf,Zwanziger:1993dh,Gribov:1977wm} as well as
with the Kugo-Ojima confinement criterion \cite{Kugo:1979gm}.

Zwanziger~\cite{Zwanziger:2003cf} has suggested that in the continuum
the behavior of both propagators in Landau gauge results from 
restricting the gauge fields to the Gribov region $\Omega$, where the 
Faddeev-Popov operator is non-negative. Generically, one gauge orbit 
has more than one intersection (Gribov copies) within the Gribov
region $\Omega$, but expectation values taken over this region are 
proposed to be equal to those over the fundamental modular
region $\Lambda$. On a finite lattice, however, this is not 
expected~\cite{Zwanziger:2003cf}. In this contribution we assess the 
importance of the Gribov ambiguity on a finite lattice for the $SU(3)$
ghost and gluon propagators as well as for the lowest eigenvalues
of the Faddeev-Popov operator.

To study these propagators in Landau gauge using lattice
simulation, all thermalized gauge field configurations $\{U_{x,\mu}\}$
have to be fixed to this gauge. On the lattice the Landau gauge
condition is implemented by searching for a gauge transformation
\mbox{${}^{g}U_{x,\mu}=g_x\, U_{x,\mu}\,g^{\dagger}_{x+\hat{\mu}}~$,}
while keeping $U_{x,\mu}$ fixed, which maximizes the functional
\begin{equation}
  F_{U}[g] \propto \sum_{x,\mu}\Re \trace \;{}^{g}U_{x,\mu}\; .
  \vspace{-1mm}
\end{equation}
This functional has many different local maxima whose number increases
as the lattice size increases or the inverse coupling $\beta$
decreases. The different gauge copies corresponding to those maxima
are called Gribov copies, due to its relation to the Gribov
ambiguity in the continuum \cite{Gribov:1977wm}.  All Gribov copies
$\{{}^{g}U\}$ belong to the gauge orbit created by $U$ and satisfy the
lattice Landau gauge condition $\partial_{\mu}{}^{g}\!\!A_{x,\mu}=0$
with
\begin{equation}\label{eq:vectorpot}
  {}^{g}\!\!A_{x+\hat{\mu}/2,\mu} = \frac{1}{2i}\left(^{g} U_{x,\mu} -
                    \ ^{g} U^{\dagger}_{x,\mu}\right)\Big|_{\rm
                    traceless}.
\end{equation}

In the literature it is widely accepted that the gluon propagator does
not depend on the choice of Gribov copy, while an impact on the
$SU(2)$ ghost propagator has been observed 
\shortcites{Bakeev:2003rr}
\cite{Cucchieri:1997dx,Bakeev:2003rr,Nakajima:2003my}. 
However, in a more recent investigation~\cite{Silva:2004bv} an
influence of Gribov copies on the $SU(3)$ gluon propagator has
been demonstrated, too.

Here we report on a combined study of the $SU(3)$ gluon and ghost propagator
in Landau gauge on the same gauge field configurations generated
at $\beta=5.8$, 6.0 and 6.2. For each configuration we
have taken $N_{cp}=30$, 40 and 10 random gauge copies for the lattice
sizes $16^4$, $24^4$ and $32^4$, respectively. A subsequent
gauge-fixing was carried out using standard over-relaxation until 
$\max_{x}(\partial_{\mu}{}^g\!\!A_{x,\mu})^2 < 10^{-14}$
was reached.

\begin{figure}[tb]
  \mbox{\hspace{1.5cm}%
    \includegraphics[width=9cm]{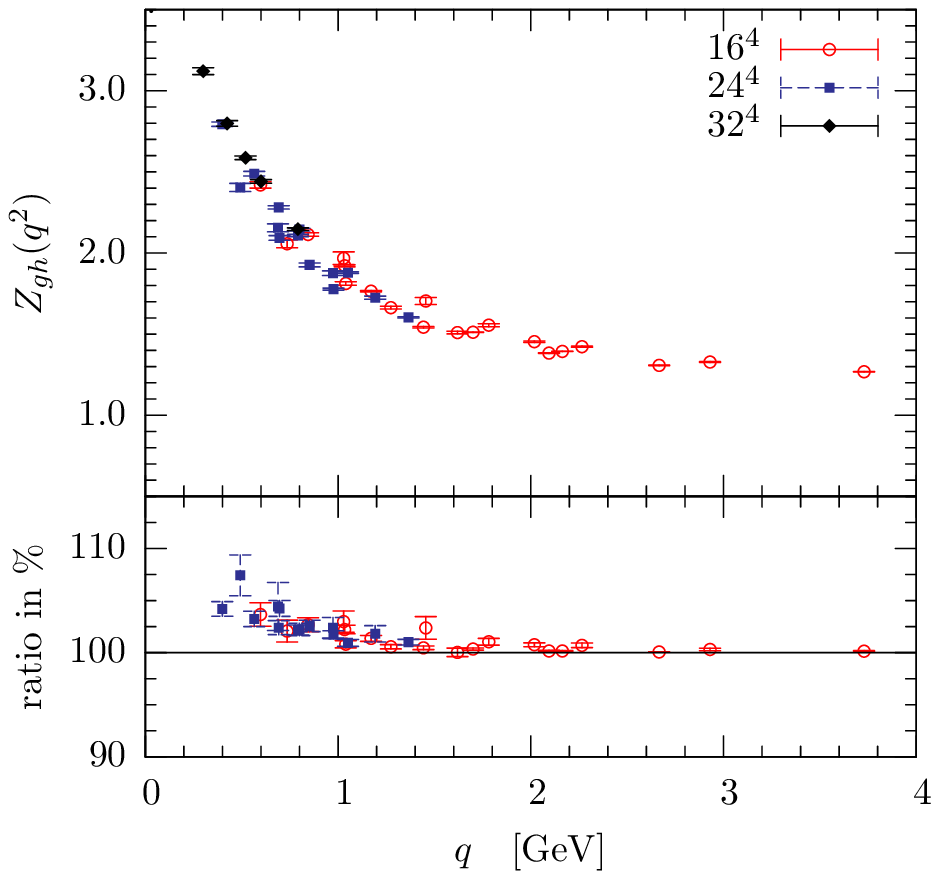}\hspace{-2.0cm}
    \includegraphics[width=9cm]{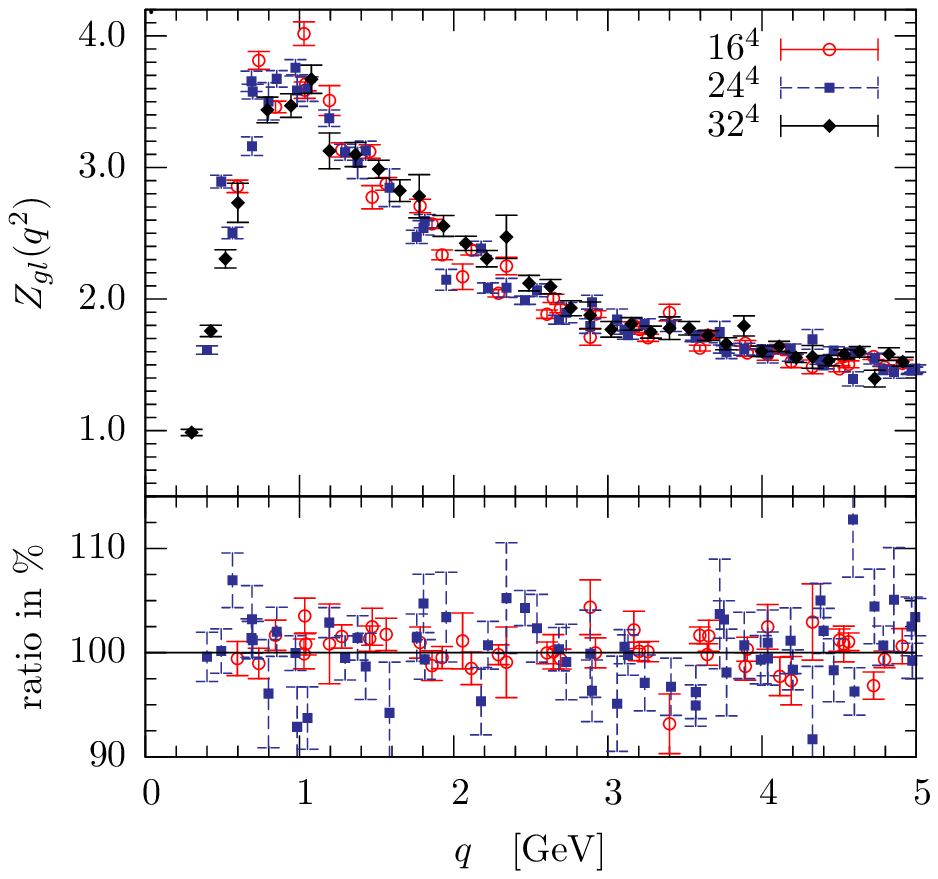}%
       } \vspace{-20mm}
  \caption{The upper parts show the dressing functions of the
           ghost $Z_{gh}$ and gluon propagator $Z_{gl}$ measured on
           best gauge copies as functions of the momentum $q$ 
           (scaled to physical units at
           $\beta=5.8, 6.0$ and $6.2$) using various lattice sizes. The
           lower parts show the ratio $\langle Z^{(\rm fc)}\rangle /
           \langle Z^{(\rm bc)}\rangle$ determined from the first (fc)
           and best (bc) gauge copies.\vspace{-0.5cm}}
     \label{fig:gh_and_gl_as_func_of_q}
\end{figure}

On each first (fc) and each best (bc) gauge copy --- that with largest
functional value among $N_{cp}$ copies --- both the ghost and the
gluon propagator have been measured. The results are shown in
\Fig{fig:gh_and_gl_as_func_of_q}. The upper parts show the dressing
functions of both propagators measured on the best gauge copies as a
function of the momentum $q$ scaled to energy units. In order to
compare to other studies~\cite{Silva:2004bv,Leinweber:1998uu} we
have used $a^{-1}=1.53$, 1.885 and 2.637 GeV for $\beta=5.8$, 6.0
and 6.2, respectively. Looking at the lower parts of this figure it
becomes clear that the ghost propagator is affected by the choice
of the Gribov copy the more the momentum is decreased. The impact 
on the gluon propagator stays inside the statistical error. For
further details we refer to \cite{Sternbeck:2004xr}.
\begin{figure}[t]
  \centering\mbox{%
    \includegraphics[width=6.8cm]{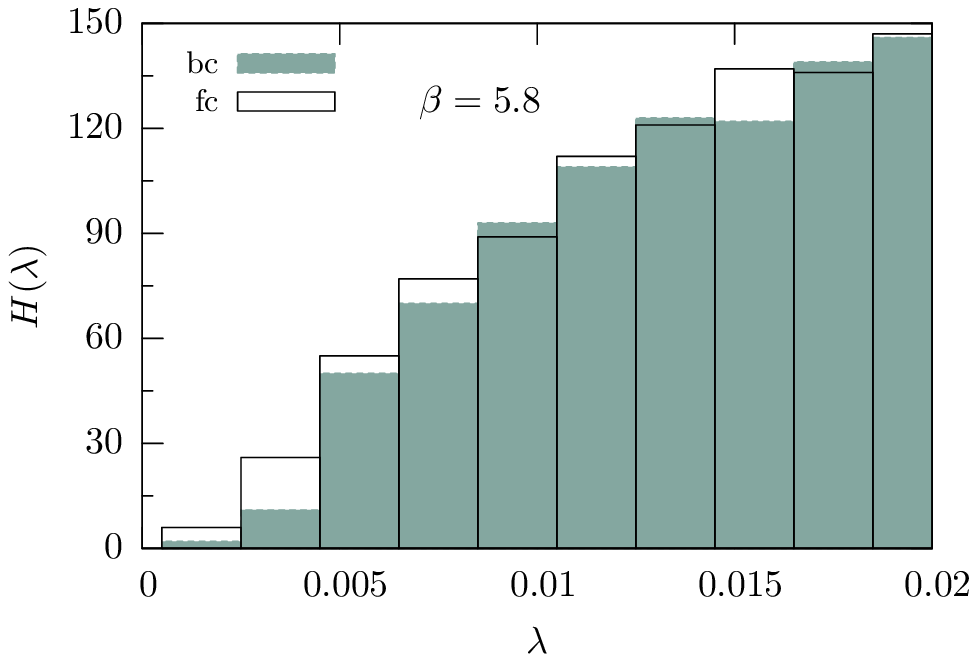}%
    \includegraphics[width=6.8cm]{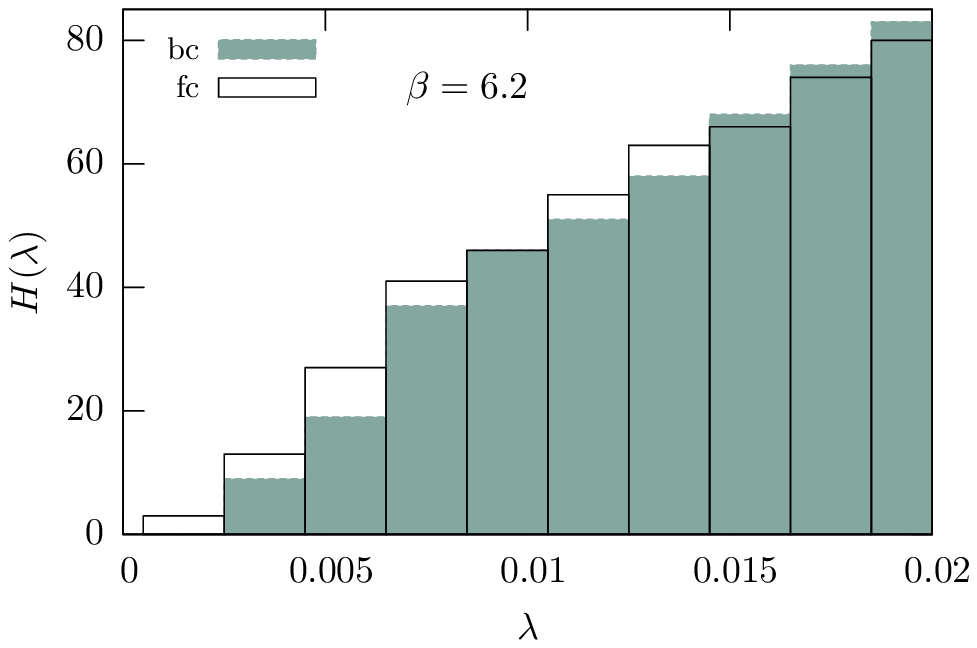}}
  \caption{The frequency $H(\lambda)$ of the lowest
    eigenvalues $\lambda$ of the Faddeev-Popov operator is shown. Full 
    boxes represent the distribution obtained on the best gauge
    copies, while empty boxes represent those on the first gauge 
    copies.\vspace{-0.5cm}}
  \label{fig:eigenvalues}
\end{figure}

In trying to fit to the proposed power laws of the dressing
function at lowest momenta, mentioned at the beginning, it turns out
the lattice sizes used are to small to confirm such a behavior.

We also calculated the eigenvalue distribution of the
Faddeev-Popov operator on the first and best gauge-fixed configurations
as shown in \Fig{fig:eigenvalues}.
Looking at this figure it is obvious that the distribution $H(\lambda)$
of the lowest lying eigenvalues~$\lambda$ on the best gauge copies is
slightly shifted towards larger eigenvalues compared to that
determined on arbitrary first gauge copies. Thus better gauge-fixing
seems to increase the gap between the lowest eigenvalues and the
Gribov horizon.


\bigskip {\small All simulations were done on the IBM pSeries 690 at HLRN.  We
  thank R.~Alkofer for discussions and H.~St\"uben for contributing
  parts of the program code. This work has been supported by the DFG
  under contract FOR~465.  A.~Sternbeck acknowledges
  support of the DFG-funded graduate school GK~271.}




\end{document}